\documentclass[runningheads]{llncs}
\usepackage{graphicx}
    \usepackage{caption}
    \usepackage{subcaption}

\usepackage{lineno}

\usepackage{amsmath}
\usepackage{multirow} 

\usepackage{datetime}

\begin{document}
%
\title{Clipping and Intersection Algorithms: \\
            Short Survey and References
    }
\titlerunning{A Survey of Line Clipping Algorithms}
%
\author{
    Vaclav Skala\inst{1}\orcidID{0000-0001-8886-4281} 
}
\authorrunning{V. Skala}
%
\institute{
    University of West Bohemia, Faculty of Applied Sciences \\
    Dept. of Computer Science and Engineering \\
    Pilsen, CZ 301 00, Czech Republic \\
    \email{skala@kiv.zcu.cz} \quad  \url{www.VaclavSkala.eu} 
    }
\maketitle              
%

\begin{abstract}
    This contribution presents a brief survey of clipping and intersection algorithms in $E^2$ and $E^3$ with a nearly complete list of relevant references.
    Some algorithms use the projective extension of the Euclidean space and $vector-vector$ operations, which supports {GPU} and {SSE} use.
    
    This survey is intended as help to researchers, students, and practitioners dealing with  intersection and clipping algorithms. 
    \footnote{
        The PDF drafts of the Skala's papers can be found at the papers repository \\
        http://afrodita.zcu.cz/$\sim$skala/publications.htm
    }

\end{abstract}
%
%


\section{Introduction}
    Intersection algorithms are the key algorithms in many areas, e.g. in geometry intersection algorithms of two lines in $E^2$  or three planes in $E^3$. 
    Many of those algorithms are part of standard courses and based on formulations in the Euclidean geometry, e.g.  Schneider-Eberly\cite{Schneider-Eberly-2003}. 
    However, there is a problem with results in an infinity or close to an infinity in some cases. 
    Some cases can be solved using the projective extension of the Euclidean space extension of duality Skala\cite{Skala2010407}. 
    The projective extension enables to represent points in an infinity and  using the principle of duality to solve  dual problems by the same algorithm Skala\cite{Skala2008615}.
    
    Also, as this approach leads to formulations using $vector-vector$ operations, it is convenient for {GPU} and {SSE} use. 
    
    Algorithms for intersections of different geometric entities in $E^2$  and $E^3$ are studied for a long time from different aspects as their robustness and the precision of numerical computations is severely influenced by the limited numerical precision available at today's computer system. It is well known, that $(1/3)*3 \ne 1$ in "the computer world". Even a simple summation $S=\sum_{i=1}^n a_i$ is not easy  Skala\cite{Skala2013SummationPR}.
    
    It should be noted that, not only in geometry oriented algorithms, a special care has to be devoted to the cases where differences between mathematics with infinite precision and mathematics with a limited precision might cause problems which could lead to the unexpected and incorrect results, sometimes also leads to disasters. 
    
    Unfortunately, programmers and computer scientists are mostly targeted to "the technology of implementation". They have a limited understanding of numerical aspects of today's numerical data representation limited more or less to the IEEE floating-point representation IEEE-754\cite{wiki:IEEE-754}. 
    Despite of the technological progress, the binary128 and binary256 precisions IEEE-754\cite{wiki:IEEE-754} are not supported in hardware.  
    
    It appears that there is no possibility to represent rational, irrational and transcendental numbers used in mathematics, where unlimited precision is expected, e.g. what is the difference of the value of $\pi ^ \pi$ and $ (\mathbf{long real}~ pi)^{(\mathbf{long real}~ pi)} $ if the IEEE-754 representation is used? 
    
    Line, half-line (ray), line segment and triangle-triangle intersection algorithms are considered as  fundamental algorithms in nearly all algorithms dealing with geometrical aspects.
   
    In computer graphics, some intersection algorithms are called clipping algorithms and serve to determine a part of one geometric entity inside of the second one. 
 

%
\section {Relevant books and journals}
    There are many books published related to intersection algorithms, clipping and computer graphics, which give more context, deeper understanding (*some books are available for free download via link.Springer.com), e.g.:
    
    \begin{itemize}
        \item Salomon,D.: The Computer Graphics Manual\cite{Salomon-2011}* \\
        \textbf{highly recommended},
        \item Salomon,D.: Computer Graphics and Geometric Modeling\cite{Salomon-1999},
        \item Agoston,M.K.: Computer Graphics and Geometric Modelling: Mathematics\cite{Agoston-2005},
        \item Agoston,M.K.: Computer Graphics and Geometric Modelling: Implementation \& Algorithms\cite{Agoston-2004},
        \item Lengyel,E.: Mathematics for 3D Game Programming and Computer Graphics\cite{Lengyel-1988}, 
        \item Vince,J.: Introduction to the Mathematics for Computer Graphics\cite{Vince-2010}* \\basic mathematical description of mathematics for undergraduates,
        \item Foley,J.D., van Dam,A., Feiner,S., Hughes,J.F.: Computer graphics - principles and practice\cite{Foley-1990},
        \item Hughes,J.F., van  Dam, A.,  McGuire,M.,  Sklar,D.F.,  Foley,J.D.,  Feiner,S.K.,  Akeley, K.:  Computer  Graphics  -  Principles  and  Practice\cite{Hughes-2014},
        \item Ferguson,R.S.: Practical Algorithms for 3D Computer Graphics\cite{Ferguson-2013},
        \item Shirley,P., Marschner,S.: Fundamentals of Computer Graphics\cite{Shirley-2009}
        \item Theoharis,T.,  Platis,N.,  Papaioannou,G.,  Patrikalakis,N.:  Graphics  and Visualization: Principles \& Algorithms\cite{Theoharis-2008}  
        \item Comninos,P.: Mathematical and Computer Programming Techniques for Computer Graphics\cite{Comninos-2005}, 
        \item Schneider,P.J., Eberly,D.H.: Geometric Tools for Computer Graphics\cite{Schneider-Eberly-2003},
        \item Ammeraal,L., Zhang,K.: Computer graphics for {Java} programmers\cite{Ammeraal20171},
        \item Vince,J.: Matrix Transforms for Computer Games and Animation\cite{Vince-2012-Transf}
    \end{itemize}        
    There are also computer graphics books using OpenGL interface, e.g.:
    \begin{itemize}
        \item Hill,F.S.,  Kelley,S.M.:  Computer  Graphics  Using  OpenGL\cite{Hill-2006},
        \item Angel,E.,  Shreiner,D.:  Interactive  Computer  Graphics\cite{Angel-2011},
        \item Hearn,D.D., Baker,M.P., Carithers,W.: Computer Graphics with OpenGL\cite{Hearn-2010} 
        \item Govil-Pai,S.: Principles of Computer Graphics: Theory and Practice Using OpenGL and Maya\cite{Govil-Pai-2005} 
    \end{itemize}
    More advanced books using Geometric Algebra and Conformal Geometry Algebra approaches are recommended for a deeper study, e.g.:
    \begin{itemize}
        \item Vince,J.: Geometric Algebra: An Algebraic System for Computer Games and Animation\cite{Vince-2009},
        \item Vince,J.: Geometric Algebra for Computer Graphics\cite{Vince-2008},
        \item Dorst,L.,  Fontijne,D.,  Mann,S.:  Geometric  Algebra  for  Computer  Science: An Object-Oriented Approach to Geometry\cite{Dorst-2009},
        \item Hildenbrand,D.:  Foundations  of  Geometric  Algebra  Computing\cite{Hildenbrand-2012},
        \item Kanatani,K.: Understanding Geometric Algebra: Hamilton, Grassmann, and Clifford  for  Computer  Vision  and  Graphics\cite{Kanatani-2015},
        \item Calvet,R.G.: Treatise of Plane Geometry through Geometric Algebra\cite{Calvet-2007}
        \item Guo,H.:  Modern  Mathematics  and  Applications  in  Computer  Graphics  and Vision\cite{Guo-2014}. \\ \\
        A short description of barycentric coordinates using geometry algebra approach is available at Skala\cite{Skala2008120}.
    \end{itemize}
    It is also recommended to study "the historical" books, e.g.:
    \begin{itemize}
        \item Newman,W.M., Sproull,R.F.: Principles of Interactive Computer Graphics\cite{Newman-1979},
        \item Harrington,S.:  Computer  Graphics:  A  Programming  Approach\cite{Harrington-1987},
        \item Mortenson,M.E.: Computer Graphics: An Introduction to the Mathematics and Geometry\cite{Mortenson-1988},
        \item Watt,A.: Fundamentals of Three-Dimensional Computer Graphics\cite{Watt-1990},
        \item Salomon,D.:   Transformations   and   Projections   in   Computer   Graphics\cite{Salomon-2006},
        \item Akenine-Moller,T.,  Haines,E.,  Hoffman,N.:  Real-Time  Rendering\cite{Akenine-Moller-2008},
        \item Eberly,D.H.: Game Physics\cite{Eberly-2003},
        \item Pharr,M.,  Jakob,W.,  Humphreys,G.:  Physically  Based  Rendering:  From Theory  to  Implementation\cite{Pharr-2016},
        \item Thomas,A.: Integrated Graphic and Computer Modelling\cite{Thomas-2008}, \\
        describes the implementation of algorithms with examples of assembler codes,
        \item Coxeter,H.S.M.,Beck,G.: The Real Projective Plane \cite{Coxeter-1992}
    \end{itemize}
    Many algorithms with codes are presented in GEMS books:
    \begin{itemize}
        \item Graphics Gems, Ed.Glassner,A.\cite{Glassner-GEMS},
        \item Graphics Gems II, Ed.Arvo,J.\cite{Arvo-GEMS-II}
        \item Graphics Gems III, Ed.Kirk,D.\cite{Kirk-GEMS-III},
        \item Graphics Gems IV, Ed.Heckbert,P.S.\cite{Heckbert-GEMS-IV}.
    \end{itemize}
    Leading computer graphics journals:
    \begin{itemize}
        \item ACM Transactions on Graphics (TOG),
        \item Computer Graphics Forum (CGF),
        \item Computers \& Graphics (C\&G),
        \item IEEE Trans. on Visualization and Computer Graphics (TVCG),
        \item The Visual Computer (TVC),
        \item Computer Animation and Virtual Worlds (CAVW),
        \item Journal of Graphics Tools (JGT),
        \item Graphical Models
    \end{itemize}
    present a wide variety of intersection algorithm applications and theories.
    
    In the following a brief classification of intersection algorithms in 2D and 3D will be made with short characteristics as Wiki\cite{wiki:Clipping}.
    The above-mentioned books present principles of the relevant methods and algorithms. 
    Also, an overview of algorithms can be found in Skala\cite{Skala-habil-1990} and the Bui's PhD\cite{Bui-PhD-1999}.
    A deeper analysis, comparisons, classifications etc. can be found in the relevant papers. 
    
    There are so many fundamental algorithms variants that differ in some aspects; mainly, the timing factor is the primary motivation.
    However, claimed speed up mostly depends on the hardware properties (caching etc.), programmer's skill and actual language and compiler used.

\section{Intersection algorithms in 2D} 
    Algorithms for intersections of different 2D geometric entities are studied for a long time from various aspects, primarily due to computation speed, robustness and limited numerical precision of the floating-point representation. The majority of 2D algorithms deals with an intersection of a line or a half-line (ray) or a line segment with 2D geometric entity, e.g. a rectangle, convex polygon, non-convex polygon, quadric and cubic curves, parametric curves and areas with quadratic arcs Skala\cite{Skala2015}\cite{Skala-EG-1989}\cite{Skala-CGI-1990}, etc.

\subsection{Intersection with a rectangular area}
    Intersection algorithms with a rectangular area (window) are well known as line clipping or as line segment clipping algorithms were developed and used for a flight simulator project led by Cohen\cite{Cohen-1969} in 1967.
    An efficient coding of a line segment position coding leading to significant computational reduction was introduced in Sproull\&Sutherland\cite{Sproull-1968} and patented by Sutherland\cite{Sutherland-clipping-patent} in 1972. 
    The Cohen\&Sutherland algorithm is described  in Newman\cite{Newman-1979}, Comninos\cite{Comninos2006}, Matthes\cite{Matthes2019-AnotherSB}, 
    Matthes\cite{Matthes-2019}, etc.

    The Cohen-Sutherland algorithm can be extended to the 3D case, i.e. intersection of a line with a cube or right parallelepiped. 
    The Cohen-Sutherland algorithm was improved by Nicholl–Lee–Nicholl\cite{Nicholl1987253} and Bui\cite{Bui199831} by classification of some possible cases.
    Ultimate cases classification was made by Skala\cite{Skala2021a1} in 2021.
    The algorithms  Liang-Barsky\cite{Liang-Barsky-1983} and  Doerr\cite{Doerr1990449} are based on direct intersection computation of a line with polygon edges in the parametric form.
    Simple and robust 3D Clipping algorithms based on projective representation and homogeneous coordinates using a separation of the convex polygon vertices by the given line was presented in 
    Skala\cite{Skala2004270}\cite{Skala2005905}\cite{Skala2012}\cite{Skala2020199}.
    
    Other known proposed algorithms can be found in references Bui\cite{Bui-PhD-1999}, Andreev\cite{Andreev1991519}, 
    Bao\cite{Bao1996741}, 
    Devai\cite{Devai2005726}\cite{Devai2006131}\cite{Devai1998157}, 
    Duvalenko\cite{Duvalenko1990}\cite{Duvanenko199339}\cite{Duvalenko1996}, 
    Cai\cite{Cai20011063}, 
    Day\cite{Day1992421}\cite{Day1992241},
    Evangeline\cite{Evangeline2014}, 
    Kaijian\cite{Kaijian1990297}, 
    Kodituwakku\cite{Kodituwakku2013}, 
    Kong\cite{Kong2001}, 
    Maillot\cite{Maillot1992276}, 
    Wei\cite{Wei20131313}, 
    Slater\cite{Slater1994407}, 
    Ray\cite{Ray-2012-a}\cite{Ray-2012-b}\cite{Ray-2014}\cite{Ray-2015}, 
    Li\cite{Li-Honglin-2016}, 
    Singh\cite{Singh-2016}, 
    Dev\cite{Dev-2019}, 
    Tran\cite{Tran-1986}, 
    Wang\cite{Wang2016702}.

    Some additional modifications of algorithms were published in 
    Brackenbury\cite{Brackenbury1984549}, 
    Chen\cite{Chen200915}\cite{Cheng-parallel-1989}, 
    Brackenbury\cite{Brackenbury1984549}, 
    Dimri2015\cite{Dimri2015}\cite{Dimri-2022}, 
    Ellriki\cite{Elliriki2019798}, 
    Hattab\cite{Hattab2014371}, 
    Iraji\cite{Iraji2011}, 
    Jiang\cite{Jiang2013157}, 
    Jianrong\cite{Jianrong2006ANA}, 
    Kumar\cite{Kumar20111}, 
    Kuzmin\cite{Kuzmin1995275}, 
    Li\cite{Li2014575}\cite{LI-Zhulin-2012}, 
    Meriaux\cite{Meriaux1984389}, 
    Molla\cite{Molla2003},
    Nisha\cite{Nisha2017a}, 
    Nisha\cite{Nisha2017b}, 
    Sobkow\cite{Sobkow1987459}, 
    Sharma\cite{Sharma1993225}, 
    Skala\cite{Skala-2014-CLIP-SIGGRAPH}, 
    Wang\cite{Wang1998410}\cite{Wang1998728}\cite{Wang201272}\cite{Wang2001}, Yang\cite{Yang198873},  Pandey\cite{Pandey-2013}  
    and 
    Bhuiyan\cite{Bhuiyan200922}.    \\
    Algorithms for clipping Bezier and Hermit parametric curve by a rectangular area was published in Skala\cite{Skala-Bezier-Clip}.
    Dawod\cite{Dawod2011} presented FPGA solution for a line clipping.
    

    Analysis and comparisons of some clipping algorithms were published in  Krammer\cite{Krammer1992253}, Skala\cite{Skala2000c}\cite{Skala1995}, 
    Nisha\cite{Nisha2017b}\cite{Nisha2017a} and 
    Ray\cite{Ray-2012-b}. 
    Algorithms for triangle-triangle intersections were presented by Sabharwal\cite{Sabharwal-2013}\cite{Sabharwal-2015}\cite{Sabharwal-2016}.

\subsection{Intersection with polygons}
    Generic solutions for polygon clipping was developed by 
    Weiler\&Atherton\cite{Weiler1977214}, Rappaport\cite{Rappoport199119},  
    Vatti\cite{Vatti1992}, Wu\cite{Wu2004228}, Xie\cite{Xie2010},  Zhang\cite{Zhang2002796}.
    Boolean operations with polygons were introduced by 
    Rivero\cite{Rivero2000881} and Martinez\cite{Martinez20091177}.
 
    Algorithms for line clipping by a polygon depend on the polygon property, i.e. if the polygon is convex or non-convex.
    In the case of convex polygons, the convexity property and ordering of vertices enable to decrease complexity to $O(lgN)$. However, in the non-convex polygon cases, when the polygon can be self-intersecting, etc., problems with a robustness of computation can be expected, in some cases a three-value logic is to be used, e.g. Skala\cite{Skala-EG-1989}\cite{Skala-CGI-1990}.

    \subsubsection{Convex polygons}
        The Cyrus-Beck's\cite{Cyrus197823} algorithm is the famous algorithm for line-convex polygon clipping. It is based on computation of a parameter $t$ of a line in the parametric form and edges of the given convex polygon.
        The algorithm is of $O(N)$ computational complexity. 
        The algorithm can be extended for the $E^3$ case.
        Some improvements and modifications were described by Skala\cite{Skala1993417}. 
        As the edges of the convex polygon are ordered, algorithm with the $O(lgN)$ was derived by Skala\cite{Skala1994517}. An algorithm based on space subdivision was described in Slater\cite{Slater1994407}.
        
        Another approach based on the implicit form of the given line and convex polygon vertices classification 
        was developed in Skala\cite{Skala2021a2} and modified by Konashkova\cite{Konashkova20141177}\cite{Konashkova20153097}.
        Another algorithm based S-Clip was described in Skala\cite{Skala2021a2}. Another algorithm for a line segment clipping based on the line segment end-points evaluation with $O(N)$ complexity was described by Matthes\&Drakopoulos\cite{Matthes-Drakopoulos-2021}.
        
        The Liang-Barsky algorithm\cite{Liang19841}\cite{Liang-Barsky-1983} is based on direct intersection computation of a line with the convex polygon edges in the parametric form and has $O(N)$ computational complexity, too.
        
        The algorithm with a run-time $O(1)$ complexity using pre-computation was developed by Skala\cite{Skala1996523}\cite{Skala1996a}.  
        The algorithm was motivated by the scan-line raster conversion used recently for solving visibility in rendering.
        The memory requirements depend on the geometrical properties of the given convex polygon.
        A comparison of the $O(1)$ algorithm with the Cyrus-Beck algorithm is presented in Skala\cite{Skala-Leder-1996}\cite{Skala-Leder-Sup-1996}.
        
        Other related algorithms or modification of existing ones were published in:  
        Sun\cite{Sun20061799},
        Vatti\cite{Vatti1992}, 
        Wang\cite{Wang2005100}, 
        Li\cite{Li2005962}, 
        Nishita\cite{Nishita1999104}, 
        Gupta\cite{Gupta2016638},
        Wijeweera\cite{Wijeweera-2019}, 
        Raja\cite{Raja2019} and  
        Sharma\cite{Sharma2016783}\cite{Sharma199251} uses the affine transformation,
        Zheng\cite{Zheng1991195} uses a linear programming method for ray-polyhedron intersection.

    \subsubsection{Non-convex polygons}
        Probably, the first algorithm dealing with non-convex polygon clipping was published by Sutherland\&Hodgman in the Reentrant polygon clipping algorithm paper\cite{Sutherland-Hodgman1974} in 1974, followed by Weiler\&Atherton algorithm for polygon-polygon clipping
        \cite{Weiler198010}\cite{Weiler1977214},         
        Rappaport\cite{Rappoport199119}, 
        
        Intersections with arbitrary non-convex non-convex polygons were described in 
        Greiner\cite{Greiner-Hormann-1998} 
        and solutions of "the singular" (degenerated) cases were described in
        Skala\cite{Skala-EG-1989} and Foster\cite{Foster:2019:CSP}. \\  
        A robust solution of triangle-triangle intersection in $E^2$ is described in  McCoid\cite{McCoid-Gander-2022}. 
        Other algorithms or modifications are described in 
        Dimri\cite{Dimri2015}, 
        Evangeline\cite{Evangeline2014}, 
        Tang\cite{Tang2009}, 
        Lu\cite{Lu2002513}\cite{Lu2002409}
        and the affine transformations are used in      
        Huang\cite{Huang201355}\cite{Huang2009}\cite{Huang2002683}.  
        \\
        
        An algorithm that handles also arcs and uses a three-value logic to handle singular cases properly, including self-intersecting non-convex triangles, was described in Skala\cite{Skala2015}\cite{Skala-EG-1989}\cite{Skala-CGI-1990}. 
        Algorithm for circular arc was described in Van Wyk\cite{VanWyk1984383}, for overlapping areas by Li\cite{Li2012485} and for circular window in Lu\cite{Lu20021133}, Kumar\cite{Kumar2018AnEL}. 
        Wu\cite{WU2006540} presented algorithm for intersection of parabola segments against circular windows. 
        The paper Kui Liu\cite{KuiLiu2007589} and Landier\cite{Landier2017138} present algorithms for union and intersection operations with polygons.

\subsection{Clipping using homogeneous coordinates}
        Homogeneous coordinates are used in computer graphics not only for geometric transformations. 
        Sproull\cite{Sproull-1968} use them in the Clipping divider in 1968. 
        Arokiasamy\cite{Arokiasamy198999} use them with duality in 1989,
        Blinn\cite{Blinn199198}\cite{Blinn1978245} described the clipping pipeline using the projective extension of the Euclidean space.
        Nielsen\cite{Nielsen19953} described use of semi-homogeneous coordinates for clipping, more general view was published by Stolfi\cite{Stolfi-1991}. 
        
        New approach to 2D clipping based on separation of the convex polygon vertices by the given line was presented in 
        Skala\cite{Skala2004270}\cite{Skala2005905}\cite{Skala2012}\cite{Skala2020199}, Kolingerova\cite{Kolingerova-1997}. 
        The principle of duality Johnson\cite{Johnson-1996} presents duality between line clipping and pint-in-polygon problem.

\section{Intersection algorithms in 3D}
    Intersection algorithms in 3D are widely used in many applications. 
    An overview of the clipping algorithms is given in the Bui's PhD
    \cite{Bui-PhD-1999}. Intersection of a line segment with a polygon in 3D was studied in Segura\cite{Segura1998587}, the intersection of polygonal models was analyzed by 
    Melero\cite{Melero201997}.     

    Algorithms for 3D clipping were over-viewed in Skala\cite{Skala-habil-1990} and Bui's PhD\cite{Bui-PhD-1999}. 
    A comparative study of selected algorithms was presented by Kolingerova\cite{Kolingerova199496}. 
    Reliable intersection tests with geometrical objects were published by Held\cite{Held-1997}.

\subsection{Line-viewing pyramid}
        Special attention was given recently to a line clipping by a pyramid in 3D due to the perspective pyramid clipping.
        The problem was analyzed very recently Cohen\cite{Cohen-1969}, Sproull\cite{Sproull-1968}, Blinn\cite{Blinn199198}\cite{Blinn1978245}\cite{Blinn-1977}, 
        Skala\cite{Skala2000b}\cite{Skala2001236}.

\subsubsection{Convex polyhedron case}
        The Cyrus-Beck's\cite{Cyrus197823} algorithm is probably the famous algorithm for line-convex polyhedron  clipping in $E^3$. It is based on computation of a parameter $t$ of a line in the parametric form and plane of the given face of the convex polyhedron.
        The algorithm is of $O(N)$ computational complexity. 
        Rogers\cite{Rogers198582} published a general clipping algorithm in 3D in 1995.
        
        The algorithm with $O_{exp}(\sqrt{N})$ complexity was described in Skala\cite{Skala1997209}.
        The algorithm is based on two planes representing the given line in $E^3$ and the search of the neighbours in the triangular mesh of the given polygon. 
        The algorithm was modified by Konashkova\cite{Konashkova20153097}.
        An interesting approach using the vertex connection table was published in Konashkova\cite{Konashkova20141177}.
        
        Using pre-computation, the algorithm in $E^3$ with a run-time $O(1)$ complexity was developed by Skala\cite{Skala1996b}; 
        a comparison was presented in
        Skala\cite{Skala-Leder-Sup-1996}.


\subsubsection{Ray-convex polyhedron}
        Extensive list of relevant publications can be found via Wiki\cite{wiki:Ray-tracing}. 
      
        Intersection of a line or a ray with a triangle can be also found in
        an algorithms presented by Havel\cite{Havel2010434} using the SSE4 instructions, 
        Xiao\cite{Xiao2020} using GPUs and Skala\cite{Skala2010407} using the barycentric coordinates computation in the homogeneous coordinates, 
        Rajan\cite{Rajan202072} uses dual-precision fixed-point arithmetic for low-power ray-triangle intersections. 
        Platis\&Theoharis\cite{Platis-Theoharis-2003} published algorithm for ray-tetrahedron intersection using the Plucker coordinates.  
        The intersection with the {AABBox} is described in Eisemann\cite{Eisemann-2007}, Kodituwakku\cite{Kodituwakku-2012}, Maonica\cite{Maonica-2017} and Mahovsky\cite{Mahovsky-2004}.
        Other algorithms are available in Sharma\cite{Sharma1993225}, Skala\cite{Skala199661}, Williams\cite{Williams-2005}, Llanas\cite{Llanas2012533}, Lagae\cite{Lagae-Dutre-2005} and Amanatides\cite{Amanatides1995RayTT}.

\subsubsection{Intersection with complex objects }
        Intersection computation with implicitly defined objects were published by 
        Petrie\cite{Petrie2020} (Real Time Ray Tracing of Analytic and Implicit Surfaces),
        intersection with a torus was published by 
        Cychosz\cite{Cychosz-1991} and alternative formulations were given in  Skala\cite{Skala2013288}.
        Intersection with general quadrics using the homogeneous coordinates  was described in Skala\cite{Skala2015} and clipping by a spherical window was published by Deng\cite{Deng2006}. 


However, as polygonal models are mostly formed by triangular surfaces a special attention is also targeted to $triangle-triangle$ intersections.

\subsection*{Triangle-Triangle intersection in 3D}
    Computation of intersection of triangles is probably the most important as nearly all Computer Aided Design (CAD) systems depend on efficient, robust and reliable computation. In the CAD systems two different data sets are usually used:
    \begin{itemize}
        \item set of triangles - there is no connection between triangles; typical example is the STL format for 3D print
        \item triangular mesh - there is information on the neighbors of the given triangles and triangles sharing the given vertex; a typical example is the winged edge or the half-edge data structures
    \end{itemize}
    An efficient triangle-triangle intersection algorithm was developed by
    Moeller-Trumbore\cite{Moeller-Trumbore-1997}. 
    Other methods or approaches were described by 
    Chang\cite{Chang2009}, 
    Danaei\cite{Danaei2017230}, 
    Devillers\cite{Devillers-2002}, 
    Elsheikh\cite{Elsheikh2014143},             
    Guigue-Devillers\cite{Guigue-Devillers-2003},
    Jokanovi\cite{Jokanovic20191347}, 
    McLaurin\cite{McLaurin-Marcum-2013},
    Roy\cite{Roy1998399}, 
    Lo\cite{Lo2004AFR}, 
    Shen\&Ann\cite{Shen-Ann-2003}, 
    Tropp\&Tal\cite{Tropp-Tal-2006}, 
    Wei\cite{Wei2014553}, 
    Ye\cite{Ye20152689}. 
    Clipping triangular strips using homogeneous coordinates was described by Maillot\cite{Maillot1991219} in {GEM II}\cite{Arvo-GEMS-II}. 
    Parallel exact algorithm for the intersection of large {3D} triangular meshes was described in {de Magalhães}\cite{DEMAGALHAES-2020}.
    Comparison of triangle-triangle tests on {GPU} was described in Xiao\cite{Xiao2020}.


\section{Conclusion}
    This contribution briefly summarizes known clipping algorithms with some extents to the intersection and ray-tracing algorithms. 
    The list of published papers related to clipping algorithms should be complete to the author's knowledge  and extensive search via Web of Science, Scopus, Research Gate and WEB search with the related topics. 
    The relevant {DOI}s were included, if found. 
    If other source was found, the relevant {URL} was included.
    Unfortunately, some papers are not publicly available. 
    
    There is a hope, that this summary will help researchers, students and software developers to find the relevant papers easily. 
    
    However, users are urged to consider a limited precision of the floating-point representation\cite{wiki:IEEE-754} and numerical robustness issues to proper handling near $singular$ cases in the actual implementations.
    
    Surprisingly, during the study and this summary preparation, there are still some problems to be explored more deeply, like a robust and efficient intersection of triangular meshes; application of triangle-triangle intersection algorithms leads to inconsistencies, inefficiency, and unreliability in general 
    Those topics will be explored more in future work.

\section*{Acknowledgments}
    The author would like to thank colleagues and students at the University of West Bohemia in Plzen, VSB-Technical University and Ostrava University in Ostrava for their comments and recommendations.
    
    Thanks also belong to several authors of recently published relevant papers for sharing their views and hints provided.
\clearpage


%
%
%
%


\bibliographystyle{splncs04}
\interlinepenalty=10000

\bibliography{Clipping-BIB.bbl}

\end{document}